\documentclass{article}

\usepackage{natbib,stfloats}
\usepackage{mathrsfs}
\usepackage{tabularx}
\usepackage{graphicx}
\usepackage[dvipsnames]{xcolor}
\usepackage{hyperref}
\usepackage{doi}
\usepackage{authblk}
\usepackage{fancyhdr}

\def\newblock{\hskip .11em plus .33em minus .07em}

\fancypagestyle{plain}{%
   \fancyhf{}   
   \fancyfoot[C]{\tiny Preprint submitted to and published as a revised full article in \textit{International   Journal of Metadata, \\ Semantics and Ontologies, 2020 Vol.14 No.1, pp.26 - 38}, DOI: \url{https://dx.doi.org/10.1504/IJMSO.2020.107792}}
   
}



\begin{document}%

\title{EngMeta -- Metadata for\\Computational Engineering}

\author[1]{Bj\"orn Schembera}
\author[2]{Dorothea Iglezakis}

\affil[1]{\small{High-Performance Computing Center Stuttgart,  University of Stuttgart, Nobelstr.~19, 70569 Stuttgart, \href{mailto:schembera@hlrs.de}{schembera@hlrs.de}}}
\affil[2]{University Library, University of Stuttgart, Holzgartenstr. 16, 70174 Stuttgart, \href{mailto:dorothea.iglezakis@ub.uni-stuttgart.de}{dorothea.iglezakis@ub.uni-stuttgart.de}}
\date{}

\maketitle

\begin{abstract}

Computational engineering generates knowledge through the analysis and interpretation of research data, which is produced by computer simulation. Supercomputers produce huge amounts of research data. To address a research question, a lot of simulations are run over a large parameter space. Therefore, handling this data and keeping an overview becomes a challenge. Data documentation is mostly handled by file and folder names in inflexible file systems, making it almost impossible for data to be findable, accessible, interopable and hence reusable. To enable and improve a structured documentation of research data from computational engineering, we developed EngMeta as a metadata model. We built this model by incorporating existing standards for general descriptive and technical information and adding metadata fields for discipline-specific information like the components and parameters of the simulated target system and information about the research process like the used methods, software and computational environment. EngMeta functions, in practical use, as the descriptive core for an institutional repository. In order to reduce the burden of description on scientists, we have developed an approach for automatically extracting metadata information from the output and log files of computer simulations. 
Through a qualitative analysis, we show that EngMeta fulfills the criteria of a good metadata model. Through a quantitative survey, we can show that it meets the needs of engineering scientists. 

\end{abstract}

\pagestyle{empty}

 \section{Introduction}

\begin{figure}
	\centering
  	\includegraphics[width=\linewidth]{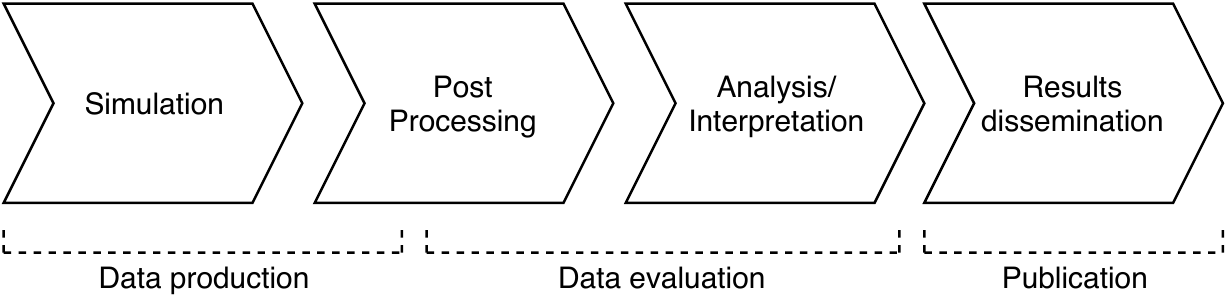}
	\caption{Scientific workflow in computational engineering.}
	\label{fig:HPCworkflow}
\end{figure}

The aim of computational engineering is the analysis of engineering problems with the help of numerical simulations. As an example of molecular dynamics (one field of computational engineering), a model of the trajectories of molecular systems can be used for the simulation of nanotubes, which have various application areas, e.g. medical applications. These simulations are performed on big computing systems, supercomputers or clusters. 

Typically, the scientific workflow in computational engineering is as depicted in figure~\ref{fig:HPCworkflow}: In the first phase of \textit{data production}, the simulation run is defined in a job file, submitted to a scheduler and then executed on the compute nodes. During the runtime of the simulation, lots of data and additional output files (such as log-files) are written to the parallel file system. After the production phase, the raw data produced in one or several simulations is prepared and analyzed in the \textit{data evaluation} phase. When the analysis of the data is completed, the researchers interpret the results by visualizing and drawing conclusions from them, resulting in the last phase, where the results are disseminated through scientific papers, which is the \textit{publication} phase.
Not every simulation results directly in a scientific publication. The massive increase in computational power has led to additional accuracy but also additional complexity of the simulations. Much more parameters have to be tested and a lot of simulations are run merely to test the computational set-up. Therefore, there is an increasing demand to manage the associated data.

During the whole workflow, the shape of research data management is quite poor~~\cite{Schembera2017}. Especially good data documentation is missing, which allows the produced research data to be made FAIR~~\cite{Wilkinson2016} as is the overall goal of research data management. Even though a lot of metadata models exist in general, none of them is suitable for the use-case of computational engineering, which is why we developed EngMeta~~\cite{Schembera2019} as a tailored model of description for this area. The development was a joint effort of the University Library of Stuttgart and the High Performance Computing Center Stuttgart together with the Institute of Thermodynamics and Thermal Process Engineering and the Institute of Aerodynamics and Gas Dynamics of the University of Stuttgart. Since the first version of the metadata model, many refinements have been made. The model is implemented in a data repository and the structured result of an automated metadata extraction of simulation files. These three topics -- an extended view on EngMeta, the automated metadata extraction and the integration in a repository -- are covered in this paper together with a qualitative and quantitive evaluation.

\section{Requirements and Related Work for a Metadata Model for Engineering Applications}



%
%
%


A relevant data description needs to include the features that allow the data to be findable, understandable and replicable in a discipline-specific context.  

For finding the data, metadata has to include information beyond the classic standard descriptive metadata.  Discipline-specific search criteria need to be included, such as information about  the target system of the simulation,  the  variables, parameters and methods used, as well as  the spatial or temporal resolution.

Understanding and hence reusing the data is a matter of information on the used software, the computational environment as well as on the encoding and on the format. Only with this information included in the data documentation, can a researcher fully grasp what has been done and how the results of the research have been produced. This is an important requirement in terms of making science reproducible, and therefore making it transparent.

For the replication of the data, data provenance is an important facet of the metadata. Information about every processing step has to be included and technical metadata has to be added.

There are existing metadata models and standards for data in general like DataCite~\cite{datacite40}, the schema underlying the metadata for DOIs or vocabularies like the W3C recommendation DCAT~\cite{dcat14}. In these standards there are description categories for citation data (title, author, publisher, dates), for subject indexing (subject, keywords) and for usage information (rights, licence, data type). However, these standards do not address the specifics of the engineering domain.

Discipline-specific models for computational engineering are hard to find.  The Chemical Markup Language CML\footnote{\url{https://www.xml-cml.org/}, last checked June 5th 2019.} and especially its extension CMLComp\footnote{\url{http://homepages.see.leeds.ac.uk/~earawa/CMLComp/index.html}, last checked June 6th 2019.} offers one approach for computational chemistry and the simulation of molecules at the atomic scale. Even though some relevant elements, such as the computational environment, are captured by the model, it is far too specific to computational chemistry. The Molecular Simulation Markup Language (MSML)~\cite{msml14} builds on CML but with an extended focus on molecular simulations. This metadata description is embedded into the MoSGrid system and serves as a workflow definition language for running the simulation and describing the outputs. Hence, a lot of manual work is involved, and the description is both specific to the workflow system and to molecular simulation.

What is missing in all these schemes are discipline-specific descriptions of the observed system and parameters of the observation itself as well as a possibility to track the provenance of the data with all relevant methods, utilities and parameters. 
\newpage
There are disciplines with elaborated and accepted metadata standards like the DDI standard for the social sciences~\cite{green2013ddi} and the CERA-2.5~\cite{CERA-2} scheme as a data-centric metadata model for climate research. The CERA model originates from 1998, being characteristic of the early commitment of the climate sciences for research data management and data description. The model incorporates discipline-specific metadata, such as the coverage of a climate phenomenon, with descriptive and process information.


\section{The Metadata Model as a Core}

Our process of deriving the metadata model started with considerations on the requirements of engineering researchers~~\cite{Iglezakis2018}: What information is important when trying to find, understand and replicate engineering data? In the next steps, we built an object model to represent this information. 

\subsection{Object Model} 
\begin{figure}
	\centering
  	\includegraphics[width=\linewidth]{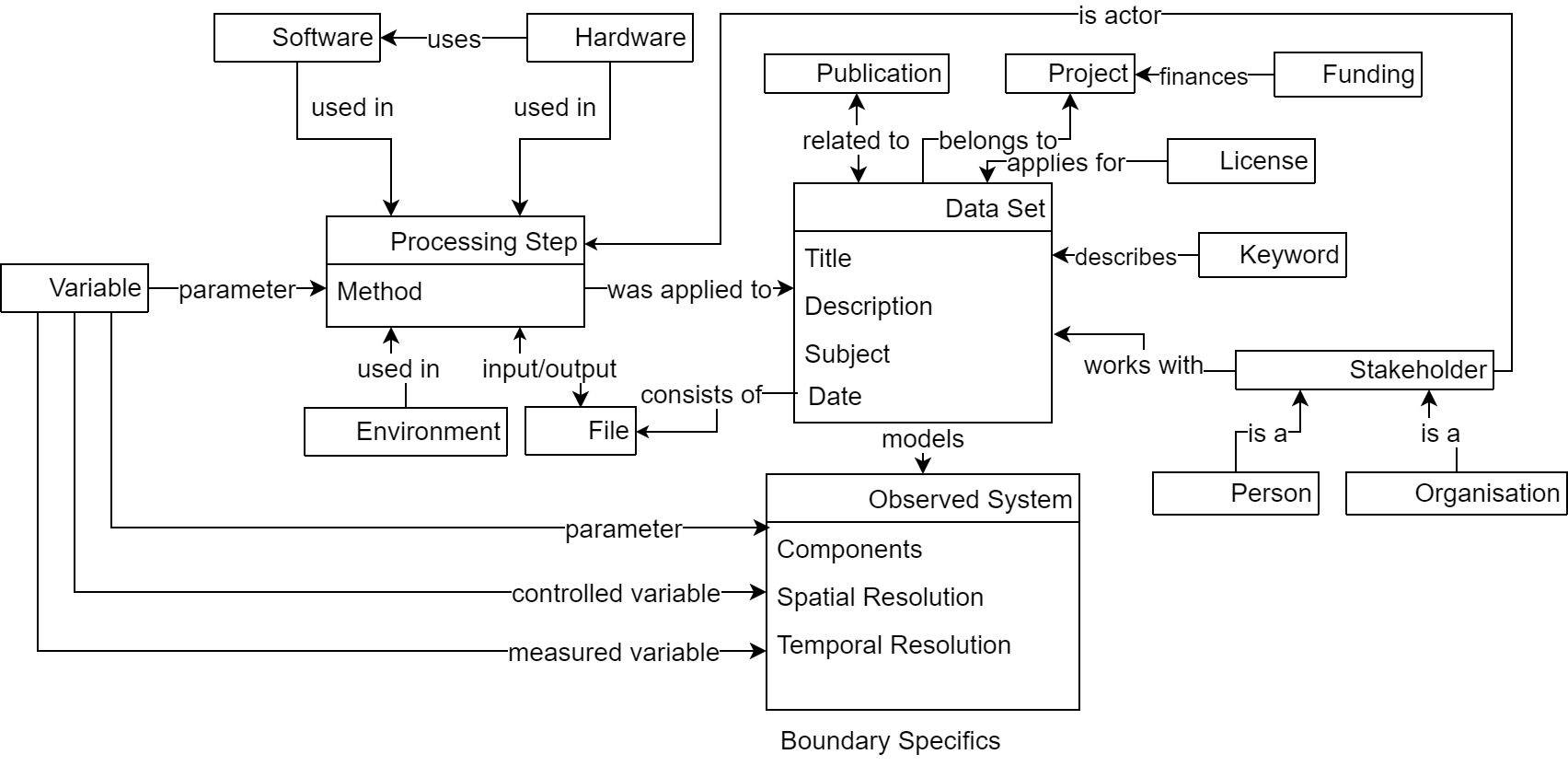}
	\caption{Model of the Objects underlying EngMeta}
	\label{fig:ObjectModel}
\end{figure}

The object model, depicted in figure~\ref{fig:ObjectModel}, is the very first step of developing a metadata model. It builds a common ground of understanding for all the relevant objects to incorporate into the metadata model. We developed this object model with the researchers from computational engineering by analyzing their scientific workflow. Even though the researchers came from distinct subject areas, they both use computer simulations to investigate their research questions. Certain entities are relevant for both subjects and are representative for the whole of computational engineering research. 

In simulation science, after the simulation run is finished, a \textit{dataset} is created  and marks the result data. The data set represents the simulated target system or \textit{observed system} as an entity in the object model. The observed system is usually characterized by controlled and measured \textit{variables} and \textit{parameters}, consists of \textit{components} and is defined by \textit{boundary conditions}. In thermodynamics, the components are molecules with \textit{force fields} acting as relevant entities in the description of research data. The observation or simulation itself has a \textit{temporal} and \textit{spatial resolution}. Moreover, the simulation \textit{method} is important to researchers for the understanding of the data. These entities form the \textbf{discipline-specific metadata}. 

The data set has been generated or processed in a \textit{processing step}. A processing step represents, for example, a simulation run, a post processing step or an analysis. The step is done within a computational \textit{environment}. This environment entity may hold information on the used hardware. Moreover, the \textit{software} used for the processing step builds its own entity, since it is critical for understanding the conducted research. The information from these entities together with the actors of the steps constitute the \textbf{process metadata}.

The data set may consist of multiple \textit{files} with file attributes like name and type and may be equipped with a \textit{PID} and a \textit{checksum} per file. These entities compose the \textbf{technical metadata}.

Moreover, the additional entities describe the data set from a descriptive point of view. These entities include related \textit{publications}, relations to other data objects representing the \textit{context}, a \textit{funding reference}, the related \textit{project}, a \textit{license}, a \textit{title}, a \textit{date}, \textit{keywords}, a \textit{description}, related \textit{persons} as well as the \textit{subject} area of the research. Moreover, a \textit{worked} entity is needed to document failed simulation runs. These entities form the \textbf{descriptive metadata}.

We decided to build a data centric metadata model, so the data set marks the central entity in the object model. The rationale behind this is that the interpretation of the data set is at the center of scientific reasoning in simulation science. When the simulation is done, the data set is analyzed with different methods resulting in new data sets. The processing step is needed as an entity to model various simulation runs or the analysis steps within one simulation project.
Although we focused on data from simulations, EngMeta is also suitable for the description of experimental data.

\subsection{The Metadata Core}
\begin{figure*}[t]
	\centering
  	\includegraphics[width=\linewidth]{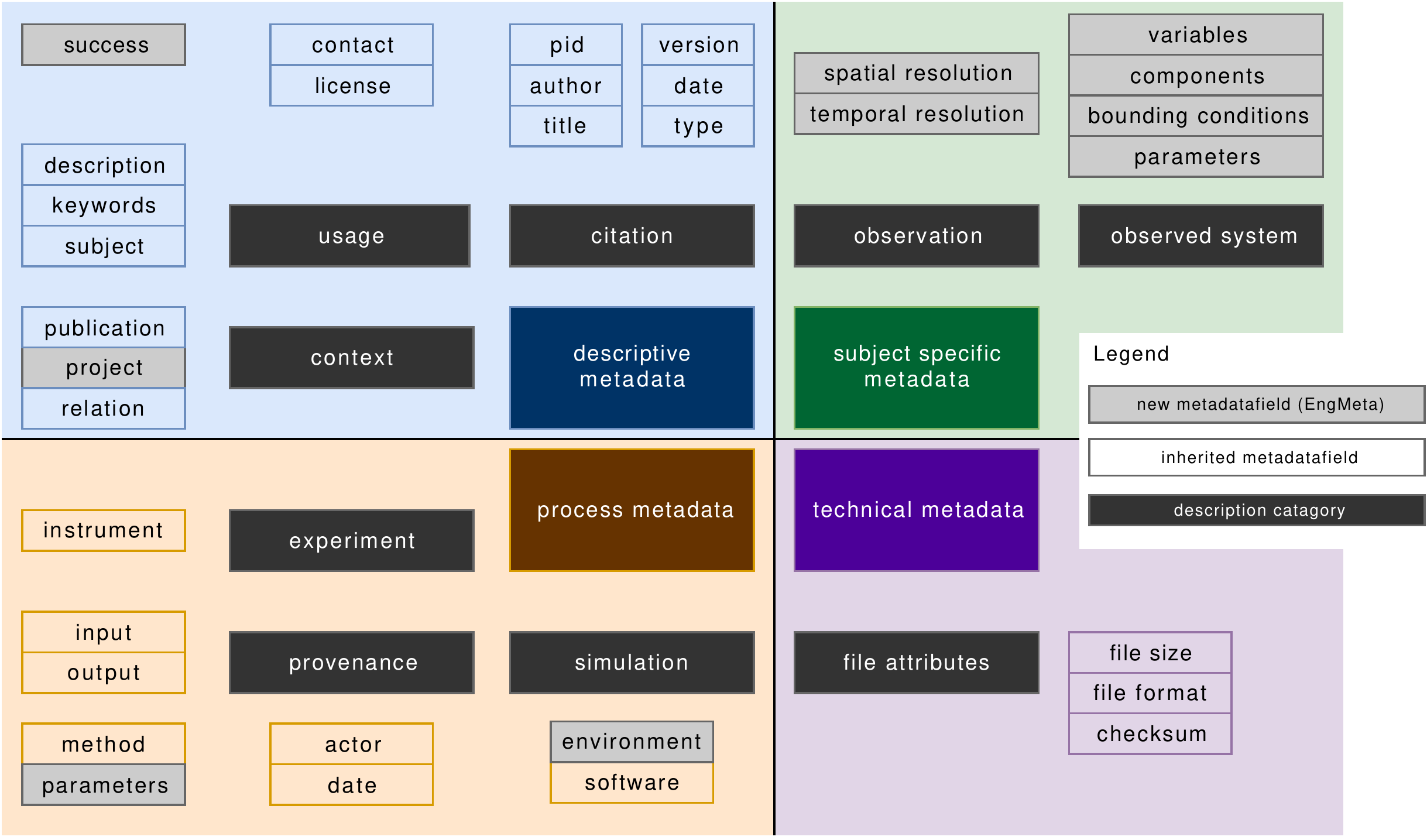}
	\caption{Components and Fields of the EngMeta Metadata Model}
	\label{fig:MetadataModel}
\end{figure*}


After being clear about the objects to be incorporated to build a relevant metadata model, we checked whether existing metadata models fit our needs. To our best knowledge, none of the existing metadata standards include all the parts of our object model. Existing metadata standards such as PREMIS or DataCite were too general, Codemeta (Software) and ExptML (Experiments) only fit for a part of the information. Discipline-specific metadata models like CML apply only for the chemical part of thermodynamics, but not for the engineering information. 
This is why we decided to build a model from scratch, building on the standards of DataCite, Codemeta, ExptML and PREMIS. We use or recommend standardized vocabularies where existing, so mainly for the general descriptive and technical metadata fields.

\textbf{CodeMeta} is the foundation for most of the metadata fields of the \textit{software} entity in our model, describing the software, such as simulation codes used to create the research data. These are the \textit{name}, \textit{contributor}, \textit{softwareVersion}, \textit{programmingLanguage}, \textit{operatingSystem}, \textit{url}, \textit{softwareSourceCode}, \textit{softwareApplication}, \textit{codeRepository}, \textit{citation}, \textit{referencePublication}. Only one element is not derived from CodeMeta.

\textbf{PREMIS}  builds the only non-CodeMeta element within the \textit{software} entity, which is the \textit{license} element. This element is formed by a \textit{pm:licenseInformation ComplexType} data type. Moreover, PREMIS is used directly inside the main \textit{data set} entity for \textit{storage} (whose data type is  \textit{pm:storageComplexType}), \textit{format} (whose data type is  \textit{pm:formatComplexType}) and \textit{rightsStatement} (whose data type is  \textit{pm:rightsStatementComplexType}).

\textbf{ExptML} is used only for the \textit{instrument} element within the \textit{processingStep} entity to describe experimental instruments used. Therefore, the \textit{ex:intrument Type} data type from ExptML is used.

\textbf{DataCite} is used throughout different metadata entities in our model. In the \textit{context} entity, which describes the related work of the research data, it is used as the \textit{relatedIdentifierType} (with the data type \textit{dtc:relatedIdentifierType}) and the \textit{relationType} (with the data type \textit{dtc:relationType}) . Within the \textit{description} entity for general descriptive information on the data object, DataCite delivers the \textit{descriptionType} element with the \textit{dtc:descriptionType}. Within the \textit{resourceType} entity, it handles the \textit{resourceTypeGeneral} element with the \textit{dtc:resourceType}. The \textit{personOrOrganization} entity, representing general information on involved stakeholders, includes the \textit{role} element as a \textit{dtc:contributor Type} DataCite type. The \textit{fundingReference} entity includes the \textit{funderIdentifierType} element, using the \textit{dtc:funderIdentifierType} data type. The \textit{title} entity uses the \textit{dtc:titleType} data type as the \textit{titleType} element. In the same way, the \textit{data} entity uses the \textit{dtc:dateType} data type as the \textit{dateType} element.
\begin{figure*}
	\centering
  	\includegraphics[width=0.7\paperwidth]{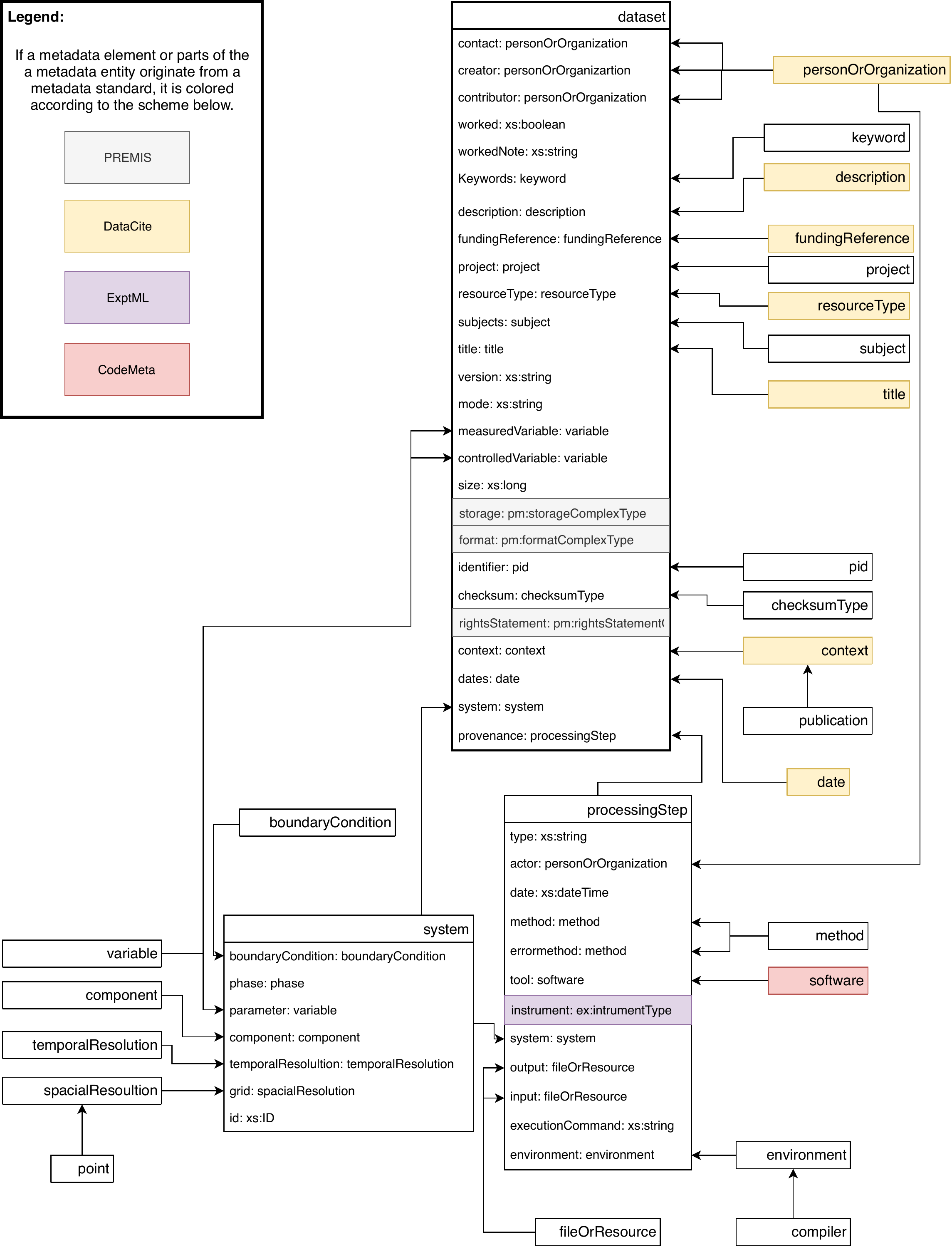}
	\caption{Metadata model with the inflated central entities \textit{dataset}, \textit{processingStep} and \textit{system} and their relation to existing metadata standards.}
	\label{fig:ObjectModel}
\end{figure*}

Figure~\ref{fig:ObjectModel} shows the metadata model with the inflated central entities \textit{dataset}, \textit{processingStep} and \textit{system} and their relation to existing metadata standards. Since the model is a data centric metadata model, we find the \textit{dataset} entity as the central entity, where all other metadata entities converge to one extensive description of a piece of research data. However, \textit{processingStep} is an important entity, since all work that is converging to a dataset is done in one or several processing steps with its parts underneath, such as the the observed system or the software used. 

The whole model is implemented as an XML schema\footnote{The full XSD file as well as an example can be found online: \url{https://bit.ly/2WQTWv3}
 , last checked on May, 24th, 2019.}, having the advantage of a strict structure that can be verified against.

Let's assume that we want to describe the results of a thermodynamical
simulation of the binding energies of two big molecules run on a HPC platform
with the help of the Open Source Software Gromacs and post-processed and analyzed with Python scripts. The components of the \textit{observed system} would be the names and SMILES codes of the molecules and the solvent with the used force field (with names and parameters as attributes) as sub-elements. The \textit{measured variable} is the distance between the molecules, the \textit{controlled variables} are the number of molecules, temperature and pressure. The \textit{temporal resolution} would be described through the number of time steps with their interval inbetween. 

There are a lot of data files produced in three processing steps. The first \textit{processing step} from type ``data generation'' describes the simulation itself. It links the input files as \textit{input} and the resulting trajectory files as \textit{output} and optionally documents the researcher as \textit{actor} and the end date of the simulation as creation \textit{date}. Gromacs would be the \textit{software} used, described with name and version and optionally with further description like a link to the source code or a describing publication. The \textit{method} would be ``thermodynamical simulation with umbrella sampling'' with the \textit{parameters} ``integrator'', ``thermostat'' and ``barostat''. The \textit{computational environment} could contain the name of the cluster, the number of nodes and cores used and optionally the compiler with its parameters.

The second \textit{processing step} from type ``post processing'' has the trajectory files as \textit{input}, the cleaned data files as \textit{output} and the python script as \textit{software} here defined by a link to the script file.

The third \textit{processing step} is from type ``analysis'' with the cleaned output file as \textit{input} and the tabular data with the summarized results as \textit{output} and the statistical method as \textit{method}s. The \textit{error method} denotes ``standard error from decorrelation'' as the information on uncertainty.

All relevant data \textit{file}s from the processing steps would be recorded with filename, link or pid and checksum.



\subsection{Crosswalk to PROV}
\label{subsec:prov}
\begin{figure*}[h]
	\centering
  	\includegraphics[width=\linewidth]{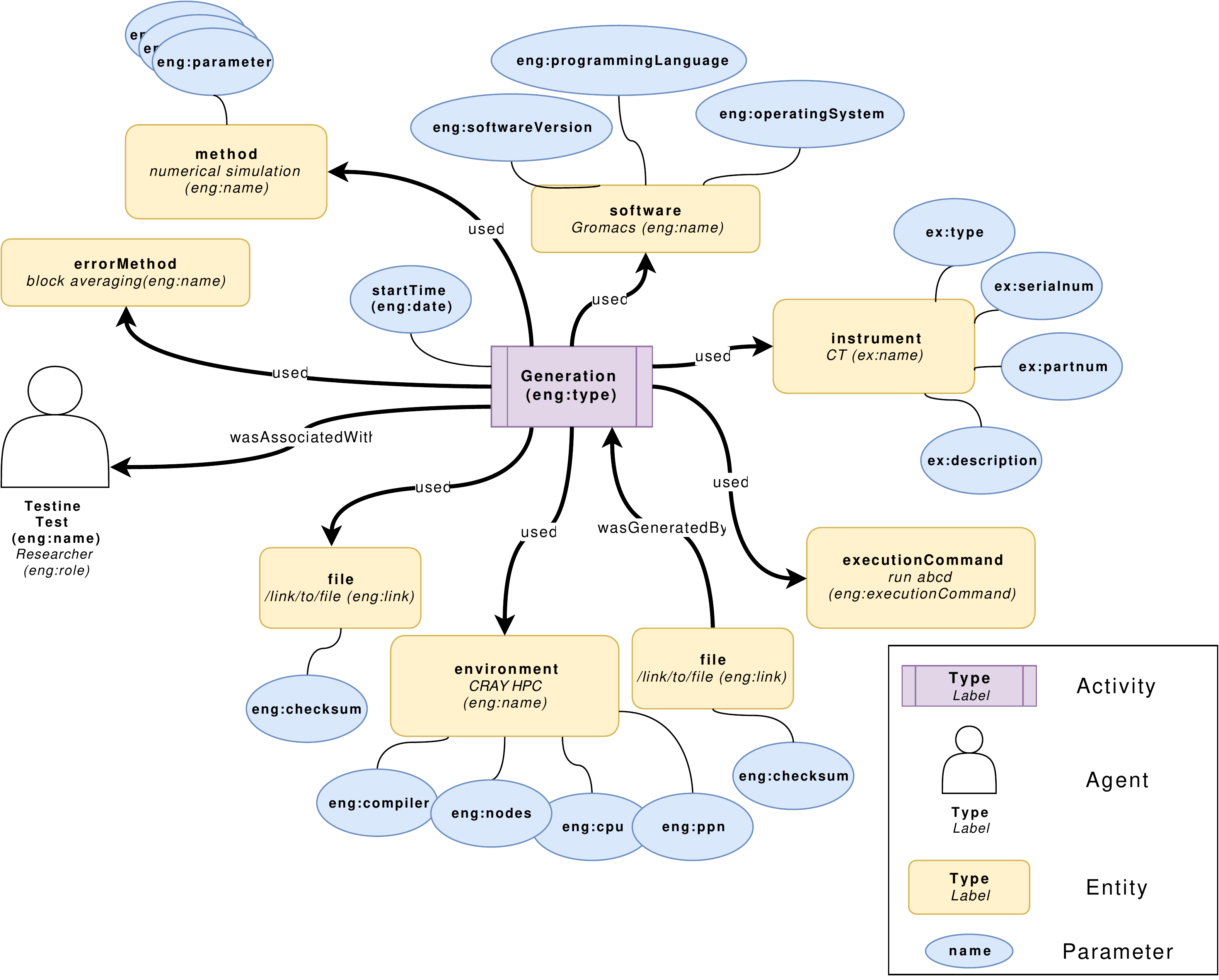}
	\caption{Conversion of a processing step of EngMeta into PROV}
	\label{fig:provEngMeta}
\end{figure*}

PROV~\citep{gil2013model} is a W3C-Standard to capture provenance information in a structured way. It comes with a data model~\citep{provdm} and with implementations in the form of an ontology and an XML scheme as well as a human readable notation. In its base model, PROV connects \textit{activities} with \textit{agents} and \textit{entities} through relations. Entities are \textit{used} or \textit{generated} by activities and \textit{attributed} to agents. Activities are \textit{associated} with agents. The provenance information of EngMeta is a list of \textit{processingSteps}. Each processingStep defines the stage in the research process (data generation, post processing, analysis, visualization), the date and actor of the step, and, optionally, input and output files, used (error)methods, software, instruments, computing environment and execution command. To convert EngMeta in PROV, each processing step becomes an activity, each actor becomes an agent and each other piece of information about a processing step becomes an entity. The activity is connected via the \textit{uses} relation with the entities for input files, methods, instruments, software, computing environment and execution command and gets the date property of the processing step to indicate the sequence of the activities. The output files are connected with the \textit{wasGeneratedBy} relation with the activity. Figure~\ref{fig:provEngMeta} visualizes the conversion of a processing step of EngMeta into PROV. 

\section{Usage of EngMeta}
EngMeta provides the opportunity to describe a dataset and its
corresponding research process with a lot of information. But the
metadata schema will only be used, if there are tools to help with the
generation and the management of the metadata. We embedded EngMeta in
a data repository based on Dataverse and implemented a tool for the
extraction of metadata already available in log and input files. 

\subsection{Automated Metadata Extraction from Simulation Files}
The tagging of metadata is a burden to the researchers. On the one hand, it is necessary for good research data management, but on the other, it is a time-consuming activity that researchers prefer to invest in their scientific endeavour. This is why we developed an approach for automated metadata extraction. In computational engineering, a lot of metadata is already available through the input, output  and log files of the simulation codes in a structured or semi-structured form. Typically, technical metadata such as filesystem attributes, process metadata such as the computational environment as well as discipline-specific metadata like controlled variables already exists in some files. For the GROMACS simulation code as an example that is used in thermodynamics, information on the computational environment is scattered through various output- and log-files of the code. An automated extraction of metadata now has the task to collect this metadata from different sources. Then, this data has to be parsed and transferred into the EngMeta scheme. The extraction is generic in a sense that the parsing of the files is directed by a configuration file. This means that the extraction is applicable to and configurable for different simulation codes and outputs and even user generated readme files.  The configuration file contains all the necessary information for parsing, which is the metadata key in terms of the EngMeta specification, the location where to search for it, the search key (how to find it), the delimiter that separates the key from the value and other information for the semantics of the results (i.e. which keys belong together). With this approach, all metadata in any textual output files of the simulation codes with syntax $<Key> <Delimiter> <Value>$ can be parsed. 

Since most supercomputer and cluster systems are based on some kind of Linux, the automated metadata extraction was implemented in Java due to interoperability. With Java, Windows operating systems are also supported. Windows support is crucial since experimental systems often operate on this operating system. We have developed two versions of automated metadata extraction. A native version, where all parsing is done linearly with the Java Scanner API and a parallel version that uses the Apache Spark Data Analytics Framework\footnote{\url{https://spark.apache.org/}, access 7.6.2019}. The rationale behind these two versions is that a native version is needed to ensure compatibility on all computer systems that support Java 1.8 and the parallel version might be needed when large output files should be analyzed and parallelisation is an advantage. However, the drawback of the latter is that one needs to have the Spark Framework installed on the cluster. 

\subsection{DaRUS Repository}
To test the metadata scheme in practice, we implemented the scheme in DaRUS, the data repository of the University of Stuttgart, based on Dataverse. Dataverse is a repository software especially for data repositories, developed by the IQSS, Harvard together with an international community of developers.  A Dataverse repository is hierarchically organized in collections, named dataverses. Each dataverse has its own user management and metadata configuration. While Dataverse is developed mainly for the publication of research data, we use it additionaly for the internal management and sharing of hot data. DaRUS therefore acts as a metadata store.  

To implement the scheme in DaRUS, we had to map the metadata fields of EngMeta with the metadata configuration of Dataverse. Dataverse comes out of the box with a set of metadata blocks: citation metadata for the general description of data and a set of discipline-specific metadata blocks for the geo sciences, social sciences, astronomy and astrophysics and the life sciences. 
Each metadata block consists of either simple fields (key-value pairs) or compound fields consisting of simple fields. New metadata schemes can be added by a super admin for the whole data repository. For each Dataverse (Collection) within DaRUS, the local admin can configure the visible, optional and required fields. 

As EngMeta is a hierarchically and multi layered XML-Schema the
challenge was to flatten the EngMeta keys and break them into suitable
metadata blocks. First, we divided the fields of EngMeta into general descriptive metadata fields, process metadata, and discipline-specific metadata. 

The citation metadata block of Dataverse covers most of the general descriptive metadata in EngMeta. We only added the possibility to mark negative results with the success field and a success note to this block.


The discipline-specific metadata includes the information about the observed system (variables, parameters, components) and information about the observation itself (spatial and temporal resolution). As the important parameters vary strongly both between and within engineering disciplines, all parameters can be added with a name and a value. So for example, instead of an extra metadata field for the Reynolds Number the researcher can add a parameter with attribute name = ``Reynolds Number'' and value=$<$value$>$. 
This procedure reduces the number of metadata fields and increases the freedom of researchers. However, it also increases the risk of typos and inconsistent names and complicates the search and filter options on this field.

Especially for the process metadata part of EngMeta that assigns the methods used with their parameters, software and hardware to individual processing steps could not be mapped 1:1 in Dataverse. As DaRUS is mainly a search index helping to find the data, we decided to extract the information out of EngMeta most likely to be searched for (software, methods, hardware and parameters used) without mapping to a processing step. To maintain the information about the process and its chronological order, the process part of EngMeta can be transformed into a PROV-File (see section~\ref{subsec:prov}) and be uploaded into the repository together with the data. 

\section{Evaluation} 

Evaluating metadata quality isn't easy, since just ``[l]ike pornography, metadata quality is difficult to define''~~\cite{Bruce2004}. However in this section, we present a qualitative evaluation based on the two frameworks  proposed in \cite{Bruce2004} and \cite{NISO2007}. Moreover, we conducted a survey among researchers where they should assess the relevance of the  EngMeta fields and check if it fits their needs.

\subsection{Qualitative Evaluation}

Here, we check EngMeta against the recommendations for good metadata as they were proposed in \cite{Bruce2004} and \cite{NISO2007}.  In the \textit{framework of guidance for building good digital collections}, six principles for metadata quality are addressed. 

The first principle for good metadata of the framework in \cite{NISO2007} refers to existing \textit{standards}. According to the first principle, existing metadata standards should be used if possible and self-built metadata schema should be avoided. Any approach should be preceded by a requirement analysis. EngMeta was designed according to this principle even though it is a ``homegrown'' scheme. We incorporated standards whereever we could, e.g. DataCite, PREMIS, ProvOne, CodeMeta and ExptML. The origin of each metadata field can be seen in figure~\ref{fig:ObjectModel}.  Moreover, we designed the first version of it in a joint effort with two engineering institutes of the University of Stuttgart, preceded by a requirement analysis published in~\cite{Iglezakis2018}. 

The second principle addresses the interoperability of the metadata scheme and means that the metadata information should be technically interoperable and understandable without knowing the context. Within EngMeta, technical or syntactical interoperability is achieved by the usage of XML as a machine-readable and system-independent format for information representation and XSD for a clear definition of the scheme. To ensure semantical understandability, the metadata model offers a wide range of attributes, being categorized into technical metadata, descriptive metadata, process-specific and discipline-specific metadata. With this information, a dataset can be understood as independent of its creators, machines and workflows.

The third principle relates to controlled vocabularies. EngMeta addresses this by using the controlled vocabularies of the incoroporated metatada standards. Moreover, some of the values for metadata entities are pre-defined by the $<xs:restriction>$ tag.

The fourth principle demands a clear statement of the terms of use. In EngMeta, this is accomplished by the $rightsStatement$ metadata field which is derived from the PREMIS metadata standard. 

The fifth principle seeks to include preservation metadata, which is definitely done by EngMeta. EngMeta supports PREMIS metadata files for long-time curation of the data, such as checksum, files sizes and file formats. Moreover, different processing steps can be defined for a dataset.

The sixth principle claims that good metadata needs to include meta-metadata, that is a description of how the metadata is structured and can be understood.  Since EngMeta is avaible as XML schema, an explanation is implicit. Additionally, the $<xs:documentation>$ tags of XML were used to give supplementary information for each metadata entity, and comments were used to give further explanation.

With respect to quality measures proposed in \cite{Bruce2004}, the first relates to completeness. This means that a metadata model should be complete in a sense that the target system is described with all the needed information. Moreover, it means that most -- or in the best case -- all elements are used later. The first part is fulfilled because EngMeta was designed with researchers from neighbouring, but not equal fields of computational engineering, so a basic common ground was determined. All entities that are included have relevance in computation engineering. This is supported by the survey we present in the succeeding section~\ref{subsec:survey}. The second part implies a quantitative evaluation. Because the repository is not yet in production, we are not yet able to present such a quantitative analysis but hope to complete this in the future.

The second criterion for a good metadata model is accuracy, meaning that it should be unambiguous and the information should be correct. This is fulfilled because EngMeta uses XML schema to have a strict definition of values, ranges, etc. Moreover, controlled vocabularies are used whereever possible.

The  third criterion aims to include provenance information, and this provenance information should also be available for the metadata itself, i.e. who created the metadata. This holds for EngMeta, since it contains possibilities to include provenance information inside the $processingStep$ entity. Multiple processing steps can be defined, where one can also be used to describe the provenance of the data creation process.

Fourth, the metadata model should conform to the expectations. This is true for EngMeta because it was developed with the computational engineering community. The survey presented in the following section~\ref{subsec:survey} supports this argument.

Fifth, metadata has to be logically consistent and coherent, meaning that the elements should be defined according to standards and standard methods should be used for metadata handling, such as crosswalks. EngMeta is a combination of the existing metadata standards DataCite, PREMIS, ExptML and CodeMeta, with a lot of additional fields that were not part of any existing model. Moreover, we implemented a crosswalk to PROV (see section~\ref{subsec:prov}). 

The sixth metadata quality measure is timeliness with respect to the link of the metadata information to the described object. In EngMeta, this is fulfilled due to the possiblity to store a PID. In our approach, in combination with the Dataverse repository, we include a DOI when uploading data and metadata to the repository. This is also the rationale for why we included DataCite as a metadata standard. 

The last quality measure refers to accessibility. This means that technical, organizational, economical and trade-related barriers should be avoided. With respect to our metadata model, the model itself is openly accessible and useable. It is understandable since XSD offers a lot of information on how to understand the metadata model. Both metadata and data described with the model can be published in the Dataverse repository, if the creator decides so. 

\subsection{Experiences}
During a test phase of the data repository, first pilot users from
different institutes of the University of Stuttgart (aerodynamics,
thermodynamics, aircraft construction, mechanics, hydraulic
engineering) tested the applicability of EngMeta for their research
data. The first results allow only first insights and no quantitative
evidence: The most frequently used parameters so far are measured
(like density or velocity) and controlled variables (like pressure and
temperature), system components and parameters (like temperature
coupling, Reynolds or Mach number) and the temporal resolution of the
simulation or observation. The controlled variables and system
parameters and components are also information the researchers want to
search and filter for. Dataverse builds on a SoLR index and offers a
full text search for textual and search facets for discrete
information, but no search interface for range queries on numerical
values. The generic definition of variables and parameters through
name and value further complicates such a numerical search.

\subsection{Survey}
\label{subsec:survey}
As the practical test of EngMeta in DaRUS only gives qualitative hints, we conducted a survey on the applicability and relevance of EngMeta for the description of research data from different engineering disciplines. The survey took place in the form of an online questionnaire in May/June 2019 at the University of Stuttgart. Five researchers took part in a pretest to determine the filling time and find any errors that may be present. The actual survey was announced at all engineering faculties of the university together with a general note in the newsletter for all employees of the University. In total, 96 researchers participated in the survey, of which 11 persons came from a non engineering discipline and were therefore excluded from the analysis, resulting in 85 participants. Most of the participants ($69\%$) denoted themselves as a researcher, $13\%$ as an institute director, $12\%$ as a group leader and one participant as a technical employee. Table~\ref{tab:surveyDisciplines} provides an overview of the disciplines of the survey participants.

\begin{table}
 \caption{Disciplines of the survey participants}\label{tab:surveyDisciplines}
 \begin{tabularx}{\columnwidth}{lX}\hline
   \textbf{Discipline} & \textbf{Percentage of Participants}\\ \hline \hline
	aerospace engineering & $31\%$ \\
	mechanical engineering & $29\%$ \\
	electrical engineering & $11\%$ \\
	civil engineering & $11\%$ \\
    process engineering & $11\%$ \\
    materials science & $7\%$ \\
    environment engineering & $7\%$ \\
    mechatronics & $6\%$ \\
    industrial engineering & $6\%$\\
\hline
  \end{tabularx}
\end{table}

$46\%$ of the participants use theoretical analysis, $71\%$ simulations and $45\%$ experiments as scientific approach. Most of the participants ($71\%$) have no experience with research data management. The datatype generated during research are mainly tabular data (by $88\%$) followed by models ($62\%$), image files ($61\%$), binary raw data ($48\%$), text files ($46\%$), software ($39\%$), video files ($24\%$), workflows ($21\%$), physical objects or samples ($19\%$) and audio files ($9\%$).

The participants estimate the relevance of the individual metadata fields of EngMeta for the description of their data on a 5-level Likert scale from $1$ to $5$. Alternatively, participants could indicate that they were unsure how to understand a metadata field. For the discipline-specific part of EngMeta we also asked for technical terms to name the individual fields and for example values of the fields. In addition to engmeta's metadata fields, we also asked participants for discipline-specific metadata categories from other areas: geo-data to specify a location or area, and information on sampling. 


\begin{figure*}[htb]
	\centering
  	\includegraphics[width=\linewidth]{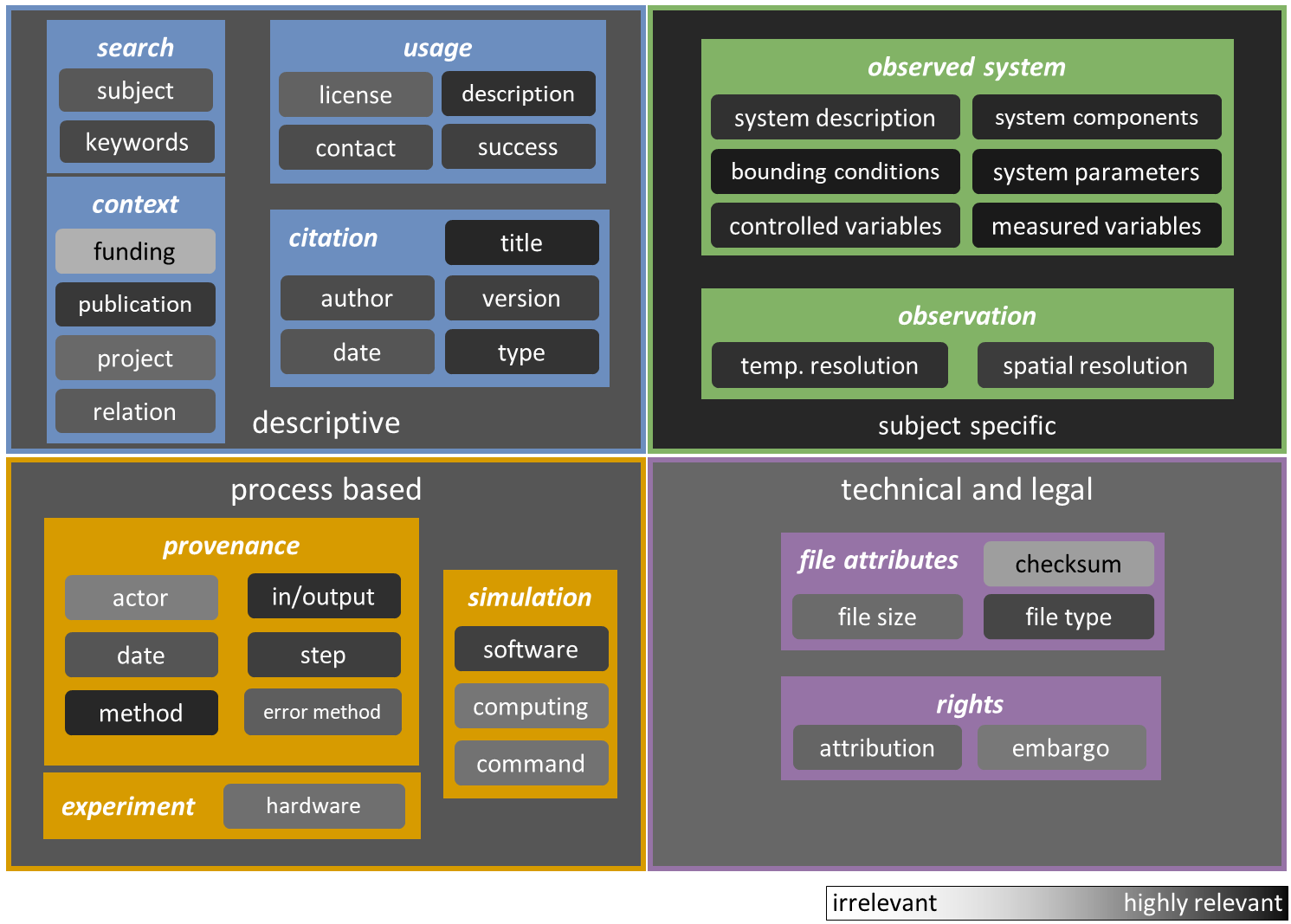}
	\caption{Mean Relevance Estimation of the Metadata Fields of EngMeta}
	\label{fig:relevance}
\end{figure*}

Figure~\ref{fig:relevance} visualizes the mean relevance estimation of the metadata fields and categories through a grey scale. The more relevant a field is for the participants, the darker is the background of the box in the drawing.

The results fit well with the experiences made within DaRUS. Most relevant for the researchers are the discipline-specific fields: the components ($m=4.39$, $s=0.81$), bounding conditions ($m=4.58$, $s=0.76$), parameters ($m=4.60$, $s=0.60$) and description ($m=4.23$, $s=0.89$) of the system, the measured ($m=4.59$, $s=0.70$) and controlled variables ($m=4.37$, $s=0.83$) and the temporal resolution ($m=4.25$, $s=1.05$) of the observation.

From the descriptive metadata, the most relevant for the engineers were title with a mean relevance of $4.40$ ($s=1.07$), description ($m=4.25$, $s=0.86$) and related publication ($m=4.12$, $s=0.93$), followed by the data type ($m=4.18$, $s=1.00$), the date ($m=4.01$, $s=1.01$) and the possibilities to mark negative results ($m=4.00$, $s=1.05$) and version the data ($m=4.00$, $s=1.01$).

For describing the research process, the participants rated highest the relevance of the used methods ($m=4.39$, $s=0.89$) (with name and parameters), input and output files ($m=4.27$, $s=0.97$), the classification of the processing step ($m=4.02$, $s=1.10$), and the used software ($m=4.00$, $s=1.02$) (specified by name and version).

From the technical metadata, the file name ($m=3.91$, $s=1.17$) and file type ($m=3.90$, $s=1.16$) were the most relevant.

There were some fields whose meaning was unclear to some of the participants. Nearly a third ($28\%$) of the participants had no notion of a persistent identifier, $20\%$ of a checksum, $12\%$ of an embargo. Interestingly, for $12\%$ of respondents, what was meant by controlled variables was also not clear. 

The least relevant were, as expected, metadata categories from another disciplines: Geo-data with a mean relevance of $m=2.18$ ($s=1.62$) and sampling with a mean relevance of $m=2.89$ ($s=1.58$). From the EngMeta fields, the least relevant information categories are funding ($m=2.22$, $s=1.24$), other collaborators apart from the authors ($m=2.87$, $s=1.11$), and technical information like the checksum ($m=2.72$, $s=2.18$). 

All other metadata fields were at least mildly relevant for the engineers with a mean relevance $> 3$.

Some of the fields were evaluated differently depending on the discipline, mostly non-scientific information like pid, licence, and embargos. But also the relevance of the spatial resolution was much more relevant for researchers from aerodynamics ($m=4.65$), civil engineering ($m=4.43$), environment engineering ($m=4.40$),  mechatronics ($m=4.25$), and mechanical engineers ($m=4.20$) than for researchers from industrial engineering ($m=3.60$), material sciences ($m=2.50$), process engineering ($m=3.33$), and electronical engineering ($m=3.50$). Information about the sampling, important in the social sciences, were highly relevant for material sciences ($m=4.50$), mildly relevant for process engineers ($m=3.67$), electronical enginneers ($m=3.57$) and civil engineers ($m=3.50$) and mostly irrelevant for scientists from aerodynamics ($m=1.92$), mechatronics ($m=2.67$), mechanical engineering ($m=2.78$) and environmental engineering ($m=2.83$). Due to the sometimes very small number of participants in some disciplines, however, these differences can only be interpreted as vague indications. 

All in all, EngMeta seems to match the information relevant and important for the researchers to describe their data. But, as the survey results imply, researchers should be relieved of the burden of dealing with information from outside the field, such as pids, checksums and legal issues, whether through automated recording or simple guidelines.

\section{Conclusions and Future Work}
EngMeta has now undergone iterative improvements and serves as the core for several efforts. For the automated metadata extraction, it defines the key to which the extracted information can be mapped. For the DaRUS research data repository, it serves as the center for data description and management. The EngMeta keys define a data object inside the repository.

EngMeta is the first attempt of a description scheme for engineering data, developed mainly with researchers from aerodynamics and thermodynamics. Whereas the evaluation suggests that EngMeta is going in the right direction, we plan to discuss the applicability and concrete structure of the fields with broader circles, both in the fields of scientists and in the field of research data management. The newly founded interest group for research data management in engineering~\footnote{\url{https://rd-alliance.org/groups/research-data-management-engineering-ig}, last checked June 6th, 2019} of the research data alliance is a good starting point on an international level, whereas the consortium for engineering in the context of the national research data infrastructure plays at the national level. In both communities, EngMeta has already been introduced. As soon as the data repository DaRUS is live, we will publish EngMeta in a proper way and register the schema in the metadata schema catalog of the RDA.\footnote{\url{https://rdamsc.bath.ac.uk/}, last checked June 6th, 2019}.

Among the most important metadata fields are variables and parameters, which can be freely specified in EngMeta with name and value and unit (and optionally with an uncertainty). This gives the scientists freedom and simplifies the scheme, but reduces the standardization and machine actionability of the content. Adding controlled vocabulary for the names and units of these fields is the next step to enhance the interoperability of the content. We are currently in dialogue with the team of the SmartCom-Project~\footnote{\url{https://www.ptb.de/empir2018/smartcom/project/overview/}, last checked June 6th, 2019} who work on such vocabularies for the engineering field.

Regarding the automated metadata extraction, the parsing as proposed in the paper works fine for basic log and output files. In the future, we will extend the metadata extraction to parse files with a syntax different from the plain $<Key> <Delimiter> <Value>$ style.

As another forthcoming work, we tend to conduct a quantitative evaluation with the metrics proposed in \cite{Gavrilis2015}. The DaRUS repository will go into production in mid 2019, so we will increase our quantitative basis for this during the year, getting more and more information about the real usage of the metadata fields.


\section*{Acknowledgement}
The DiplIng project is funded by the Federal Ministry of Education and Research under Grant No. FDM-008.


\end{document}